\documentstyle[12pt]{article}
\begin{document}

\newcommand{\Winf}{$W_{1+\infty}$ }
\newcommand{\glinf}{$\mbox{gl}(\infty)$ }
\newcommand{\glinfh}{$\hat{\mbox{gl}}(\infty)$ }
\newcommand{\be}{\begin{equation}}
\newcommand{\ee}{\end{equation}}
\newcommand{\ba}{\begin{eqnarray}}
\newcommand{\ea}{\end{eqnarray}}
\newcommand {\sfrac}[2]{{\textstyle \frac{#1}{#2}}}
\newcommand {\n}{\nonumber \\}
\newcommand{\eq}[1]{(\ref{#1})}
\newcommand{\bb}{{\bf b}}
\newcommand{\cc}{{\bf c}}
\newcommand{\bt}{{\tilde b}}
\newcommand{\vph}{{\hat{\varphi}}}
\newcommand{\qb}{{\bar q}}
\newcommand{\ket}[1]{|#1\rangle}
\newcommand{\bra}[1]{\langle #1|}
\newcommand{\fket}{{||0\gg}}
\newcommand{\fbra}{\ll 0||}
\newcommand{\flambda}{||\Lambda\gg}
\newcommand{\qed}{{\bf Q.E.D.}}
\newcommand{\vac}{\ket{\lambda}}
\newcommand{\bvac}{\bra{\lambda}}
\newcommand{\ppm}{\prod_{\epsilon=\pm 1}}
\newcommand{\ep}{\epsilon}
\newcommand{\Jc}{{\cal J}}
\newcommand{\ZZ}{{\bf Z}}
\newcommand{\CC}{{\bf C}}
\newcommand{\flatd}{\flat^\dagger}
\newcommand{\finf}{\ll-\infty ||}
\newcommand{\dt}[1]{\mbox{det}[#1]}
\newcommand{\scr}[1]{{\footnotesize #1}}
\newcommand{\vv}{{\bf v}}
\newcommand{\Eb}{{\bar E}}
\newtheorem{definition}{\bf Definition}
\newtheorem{theorem}{\bf Theorem}
\newtheorem{corollary}{\bf Corollary}
\newtheorem{lemma}{\bf Lemma}
\newtheorem{proposition}{\bf Proposition}
\newcommand{\bZ}{{\bf Z}}
\newcommand{\bC}{{\bf C}}
\newcommand{\KET}{\vert\lambda\rangle}
\newcommand{\BRA}{\langle\lambda\vert}
\newcommand{\nr}{r}%or {n} %by awata
\newcommand{\ms}{s}%or {m} %by awata
\newcommand{\rn}{n}%or {r} %by awata
%%%%%%%%%%%%%%%%%%% titlepage %%%%%%%%%%%%%%%%%%%%%%%%%%%%%%%%%%%%
\begin{titlepage}
%\topmargin -1.5 true cm
%\textheight 24.5 true cm
%\textwidth 15 true cm
%\oddsidemargin .5 true cm
%\evensidemargin .5 true cm
%
\nopagebreak
%
%
%\markright{Preliminary Draft: 94.05.12}
\begin{flushright}
May 1994\hfill
%RIMS-982\\
YITP/K-1060\\
(Revised Version: September 1994)\hfill
YITP/U-94-17\\
SULDP-1994-3\\
hep-th/9405093
\end{flushright}

\vfill
\begin{center}
{\Large
Character and Determinant Formulae}\\
\vskip 2mm
{\Large of Quasifinite Representation of the \Winf Algebra}

\vskip 20mm

{\large H.~Awata\footnote{
Research Institute for Mathematical Sciences, Kyoto University,
Kyoto 606, Japan; E-mail:awata@kurims.kyoto-u.ac.jp},
M.~Fukuma\footnote{
Yukawa Institute for Theoretical Physics,
Kyoto University, Kyoto 606, Japan; E-mail: fukuma@yukawa.kyoto-u.ac.jp},
Y.~Matsuo\footnote{Uji research center,
Yukawa Institute for Theoretical Physics,
Kyoto University, Uji 611, Japan; E-mail: yutaka@yukawa.kyoto-u.ac.jp},
S.~Odake\footnote{
Department of Physics, Faculty of Liberal Arts,
Shinshu University, Matsumoto 390, Japan;
E-mail: odake@jpnyitp.yukawa.kyoto-u.ac.jp}
}
\end{center}
\vfill

\begin{abstract}

We diagonalize the Hilbert space of some subclass of the
quasifinite module of the \Winf algebra.
States are classified according to their eigenvalues
for infinitely many commuting charges and the Young diagrams.
The parameter dependence of their norms is explicitly derived.
The full character formulae of the degenerate representations
are given as summation of the bilinear combinations of the Schur
polynomials.

\end{abstract}
\vfill
\end{titlepage}
%
%%%%%%%%%%%%%%%%%%%%%%%%%%%%%%%%%%%%%%%%%%%%%%%%%%%%%%%%%%%%%%%%
\section{Introduction}
The detailed study of (infinite dimensional) Lie algebras
has been sometimes very essential in theoretical physics.
The  representation theory of finite dimensional
Lie algebra is indispensable to understand quantum mechanics
or gauge theories.  If we extend the dimension by one,
the loop algebras such as Virasoro \cite{rBPZ}
or Kac-Moody algebras
are essential tools to describe two-dimensional statistical
systems or string theories.

Recently, in many places such as two-dimensional quantum gravity
\cite{rFKN}--\cite{rIM}, the quantum
Hall effects \cite{rCTZ}\cite{rIKS},
the membrane \cite{rBST}\cite{rFK}, or
the large $N$ QCD \cite{rGT}\cite{rDLS},
the \Winf algebra is regarded as
the fundamental symmetry of system.

As a member of loop algebras, the \Winf algebra has a unique character
in that the number of currents is infinite.
In a sense, it may be regarded as the symmetry of three-dimensional
system since it is closely connected with
the area-preserving diffeomorphism \cite{rPRS}\cite{rBK}.
Due to this fact, the detailed representation
theory was not fully developed until now
although some attempts were made \cite{rO}.
The situation is also similar in the extensions of the \Winf algebra
\cite{rFFZ}--\cite{rAFMO2}.
One of the confusing feature of the \Winf algebra
is its hybrid nature in dimensions.
We remark that
it has also definite ``two-dimensional'' aspects since we
already knew the explicit realization in terms of two-dimensional
free fields \cite{rBK}\cite{rM}.
{}Furthermore, this symmetry is found even
in instanton physics in four dimensions \cite{rT}--\cite{rYC}.

Last year, Kac and Radul \cite{rKR} discovered a way to avert
from the difficulty and proved that the Hilbert
space at each energy level can be finite dimensional
if we choose the weight vector properly.
In our previous letter \cite{rAFMO1},
we give the computer calculation of the Kac formula
of the \Winf algebra at lower degree.
In this article, we would like to give its analytical formula.
Actually, we can go further to give the explicit form of the
diagonal basis of the Hilbert space with respect
to the inner product and give their parameter dependence.
As corollaries, we give the full character
formulae \cite{rAFOQ} of any degenerate representations.
This  will be  the basis for the application of the representation
theory of the \Winf algebra to physical systems, such as quantum
gravity, the quantum Hall effects, the two-dimensional QCD
which we would like to report in our future issues.

The plan of this paper is as follows.
In section 2, we give a brief review of the
result of Kac and Radul, and also a summary of
our computer calculation of the determinant formula.
The parameters of the system can be roughly classified
into two groups, the central charges and the spins.
Our determinant formula is factorized into
functions which depend only on either of them.
%In section 3, we first derive the spin dependence
%by counting the null states which may be
%immediately predicted by the property of characteristic polynomials.
In section 3, we give the detailed
account of the relation with the \glinf algebra.
This is an essential step to understand the determinant formula.
As we see in the following sections,
the trasformation from the basis of the \Winf algebra to the
corresponding ones of the
\glinf algebra gives the spin dependent part of the
determinant formula.
On the other hand, the determinant
of the \glinf algebra explains the central charges dependence.
In section 4, we first derive the spin dependence
from this viewpoint.
%central charge dependence.
%It also gives another explanation for the spin
%dependence of the determinant formula.
In section 5, the central charge dependence is derived.
There, the knowledge of the permutation group is essential
to classifying the Hilbert space. Indeed, we derive
the explicit form of the diagonal basis with respect
to the inner product by using the Young diagrams.
In section 6, we give the character formula
for the degenerate representation  as a
bilinear form of the Schur polynomials.
In appendix A, we give tables of the determinant formula
which we previously derived by computer analysis.
In appendix B, we explain the free-fermion method
which was essential to calculating the inner product formula.
In the \Winf algebra, there are an infinite number of ``modular
parameters'' because the number of Cartan elements is infinite.
The fermion which appears here is the ``fermionization''
of those modular parameters.

%%%%%%%%%%%%%%%%%%%%%%%%%%%%%%%%%%%%%%%%%%%%%%%%%%%%%%%%%%
\section{Brief review of the \Winf algebra}
The \Winf algebra is a central extension of the
Lie algebra of the (higher order) differential operators
on the circle, which is generated by
$z^r D^k$ with $r\in\ZZ$, $k\in\ZZ_{\ge 0}$
%the polynomials of $z$
and $D\equiv z\frac{\partial}{\partial z}$.
We write the generator of the \Winf algebra which
correspond to the differential operator $z^rD^k$ as
$W(z^rD^k)$.
The commutation relations are,
\ba
  &&\left[ W(z^r f(D)), W(z^s g(D))\right]=\n
  &&
  W(z^{r+s}f(D+s)g(D))-W(z^{r+s}f(D)g(D+r)) \n
  &&+C \Psi(z^r f(D), z^s g(D)),\label{I1}
\ea
where $f(D)$ and $g(D)$ are polynomials of $D$
and we introduce the {\em two-cocycle} $\Psi$,
\ba
&&\Psi(z^r f(D), z^s g(D))=-\Psi(z^s g(D),z^r f(D))\nonumber\\
&&=\left\{
\begin{array}{ll}
        \sum_{1\leq j\leq r}f(-j)g(r-j) &
        \mbox{if  $r=-s>0$}  \\
        0 & \mbox{if  $r+s\neq 0$ or $r=s=0$}.
\end{array}
\right.
\label{I2}
\ea
The principal gradation of the \Winf algebra is
\ba
{\cal W}_{1+\infty}&=&
\bigoplus_{\nr\in\ZZ}({\cal W}_{1+\infty})_\nr\n
({\cal W}_{1+\infty})_\nr&=& \left\{
z^\nr f(D)|f(w) \in \CC[w]\right\}
\ea
It is defined in terms of the eigenvalue
of the ``energy operator''
$L_0\equiv -W(D)$.
The highest weight state of the \Winf module
is defined in terms of this gradation,
\be
\begin{array}{lcl}
        W(z^r D^k)|\lambda\rangle=0, & \qquad &
        r\geq 1, k\geq 0,
        \\
        W(D^k)|\lambda\rangle = \Delta_k |\lambda\rangle ,  &
        \qquad & k\geq 0.
\end{array}
\label{eHWC}
\ee
We introduce,
\begin{equation}
        \Delta(x)\equiv-\sum_{k=0}^\infty {x^k\over k!}\Delta_k,
        \label{I3}
\end{equation}
to rewrite the (infinite dimensional) weight vector,
which will be called the weight function.
The Verma module is spanned by the vectors which are obtained by
applying the generators of
negative gradation to the highest weight state,
$$
W(z^{-\nr_1}f_1(D))\cdots W(z^{-\nr_N}f_N(D))\vac,
$$
and we define the energy level of this state by the sum,
$\sum_{i=1}^N\nr_i$.

%A quotient of the Verma module of \Winf is called {\it quasifinite}
A representation of \Winf is called {\it quasifinite}
if and only if there are only finite number of states at each energy level.
The quasifinite module has the following properties \cite{rKR}:
\begin{enumerate}
        \item   For each level $r$ , there are infinitely many
        null generators of the form $W(z^{-r} b_r(D)g(D))$,
        where $b_r(D)$ is a monic, finite degree polynomial
        of operator $D$.

        \item   The polynomial $b_r(D)$ with $r>1$ is related to
        level-1 polynomial $b(D)\equiv b_1(D)$ as
        \begin{itemize}
                \item  $b_r(D)$ is divided
				by {l.c.m.}($b(D)$, $b(D-1)$,
                $\cdots$ $b(D-r+1)$).

                \item  $b(D)b(D-1)\cdots b(D-r+1)$ is
                divided by $b_r(D)$.
        \end{itemize}
        If the difference of any two distinct roots
		is not an integer,
        $b_r$ can be
        uniquely determined as $b_r(D)=\prod_{s=0}^{r-1}b(D-s)$.
        %Otherwise, however, the argument in \cite{rKR}
        %alone does not fix $b_r$ uniquely.

        \item The function $\Delta(x)$
        satisfies a differential equation,
        \be
        b(\sfrac{d}{dx})\left((e^x-1)\Delta(x)+C\right)=0.
        \label{eDE}
        \ee
        When $b(w)=(w-\lambda_1)^{K_1}
        \cdots(w-\lambda_\ell)^{K_\ell}$,
        the solutions are
                \be
        \Delta(x)={{\sum_{i=1}^\ell p_{K_i}(x)
		e^{\lambda_i x}-C}\over%
        {e^x-1}},~~~
        \deg p_{K_i}= K_i-1,
        \label{eDelta}
        \ee
        with $\sum_{i=1}^\ell p_{K_i}(0)=C$.
\end{enumerate}

In this article, we analyze the irreducibility of the quasifinite module.
We introduce,
%%%%%%%%%%%%%%%%%%%%%%%%%%%%%%%%%%%%%%%%%%%%%%%%%%%%%%%%%%%%%%%%%%%%
\begin{definition}{\bf Generalized Verma Module and Kac Determinant}
\newline
The second property of the quasifinite representation
means that there are at most $rK$ independent generators at
level $r$ ($W(r^{-r}D^s)$ with $s=0,1,\cdots, rK-1$)
 if the characteristic polynomial $b(w)$ has degree $K$.
We call the module freely generated by those generators
as the generalized Verma module.
The number of states at each level is given  by
the generating function \cite{rAFMO1},
\be
\prod_{r=1}^\infty \frac{1}{(1-q^r)^{r K}}\equiv
\sum_{\ell=0}^\infty n_\ell q^\ell.
\ee
At level $\ell$, we define the  determinant
of the $n_\ell \times n_\ell$ matrix which consists of
the inner products of the basis of the module
as the Kac Determinant.
\end{definition}
%%%%%%%%%%%%%%%%%%%%%%%%%%%%%%%%%%%%%%%%%%%%%%%%%%%%%%%%%%%%%%%%%%%%

The purpose of this paper is to calculate this determinant,
and with this knowledge, to give the character formula.
We restrict ourselves to  consider the special cases,
%We parametrize the characteristic polynomial
%and corresponding weight function as,
\be
b(w)=\prod_{i=1}^K (w-\lambda_i),\qquad
\Delta(x)=\sum_{i=1}^K C_i
\frac{e^{\lambda_i x}-1}{e^x-1}.
\label{2.1}
\ee
In other word, we postulate first that
there are only simple zeros in $b(w)=0$
and the difference of their roots are not integers.
%The number of independent  generators with
%energy eigenvalue $\nr$ is therefore $\nr K$.
In our previous letter \cite{rAFMO1}, we made the computer
calculation of the determinant formulae
at lower levels with this assumption.
We summarize our results in appendix A \cite{rAFMO1}.
The determinant formula for general cases
can be obtained by taking a suitable limit
of the parameters.
It may be symbolically written in the following form,
\be
\mbox{det[$\nr$]}= \prod_{i}
A_\nr(C_i)\prod_{i<j}B_\nr(\lambda_i-\lambda_j).
\label{2.2}
\ee
The functions $A_\nr$ and $B_\nr$ have zero
only when $\lambda_i-\lambda_j$ or $C_i$ is integer.
In the following sections, we derive these functions
analytically.

%%%%%%%%%%%%%%%%%%%%%%%%%%%%%%%%%%%%%%%%%%%%%%%%%%%%%%%%
%
%         Section 3 Relation with the gl-infty algebra
%
%%%%%%%%%%%%%%%%%%%%%%%%%%%%%%%%%%%%%%%%%%%%%%%%%%%%%%%
\section{Relation with the \glinf algebra}
Some of the essential features of the \Winf algebra can be
more clearly elucidated if we use the connection with
its simpler cousin, the \glinf algebra. As explained in
\cite{rKR}, we may construct a
quasifinite representation of the \glinf algebra
which is deeply connected with
the corresponding one of the \Winf algebra.
We here would like to explain the
full detail of this correspondence
since it illuminates the $\lambda$ dependence of the
determinant formula and also is essential to
calculating the $C$ dependence.

%%%%%%%%%%%%%%%%%%%%%%%%%%%%%%%%%%%%%%%%%%%%%%%%%%%%%%%%
\subsection{The \glinf algebra and its representation}
The \glinf algebra is generated by the
operators, $\Eb^{(\mu)}(i,j)$ ($\mu=0,1,\cdots,m$),
which act on the infinite dimensional
space spanned by the basis, $\vv^{(\mu)}_k$ with $k\in \ZZ$,
\be
\Eb^{(\mu)}(i,j)\vv^{(\nu)}_k=\theta(m-\mu-\nu)
\delta_{j+k,0}\vv^{(\mu+\nu)}_i.
\label{3.1}
\ee
Here $\theta(i)=1$ for $i\geq 0$ and $\theta(i)=0$
for $i<0$.
The commutation relation is,
\ba
&&\left[ \Eb^{(\mu)}(i,j), \Eb^{(\nu)}(k,\ell)\right]
\nonumber\\
&&= \theta(m-\mu-\nu)\left(
\delta_{j+k,0}\Eb^{(\mu+\nu)}(i,\ell)-
\delta_{\ell+i,0}\Eb^{(\mu+\nu)}(k,j)\right).
\label{glinf}
\ea

As usual, the highest weight state is  defined by
using the gradation,
$\mbox{deg}\>\Eb^{(\mu)}(i,j)=i+j$,
\ba
\Eb^{(\mu)}(i,j)\vac &=& 0, \qquad i+j>0,\n
\Eb^{(\mu)}(i,-i)\vac &=& \qb^{(\mu)}_i \vac.
\label{gl-hwc}
\ea
{}For the finite dimensional case, the parameters $\qb^{(\mu)}_i$
are arbitrary.  However, as the \Winf algebra,
there should be severe constraint on them once
we require  the quasifiniteness.

Define,
\be
h^{(\mu)}_k\equiv \qb^{(\mu)}_k -\qb^{(\mu)}_{k-1}.
\label{step}
\ee
We introduce the set,
$$
S^{(\mu)}\equiv \left\{
k\,|\, h^{(\nu)}_k\neq 0\mbox{ for some }\nu\geq \mu \right\},
$$
which satisfies the inclusion relation,
\be
S^{(0)}\supseteq S^{(1)} \supseteq \cdots \supseteq S^{(m)}.
\label{incl}
\ee
A quasifinite representation is then obtained \cite{rKR}
if and only if $S^{(\mu)}$ is a finite set for each $\mu$.

Let us count the non-vanishing elements in the Hilbert
space.  At level 1, the Hilbert space consists of the vectors
of the form, $\Eb^{(\mu)}(k-1,-k)\vac$.  To see if they are null,
we compute,
$$
\Eb^{(\nu)}(k,-k+1)\Eb^{(\mu)}(k-1,-k)\vac=
\theta(m-\mu-\nu)h^{(\mu+\nu)}_k\vac.
$$
It shows that it becomes non-vanishing only if
$k\in S^{(\mu)}$.
Similar computation shows that more general
state $\Eb^{(\mu)}(\ell,-k)\vac$ becomes non-vanishing only if
there exist an integer $s\in S^{(\mu)}$ such that $k\geq s> \ell$.
In the figure below, we show the
elements which become non-vanishing
for this case.

%%%%%%%%%%%%%%%%%%%%%%%
%
%  Figure
%
%%%%%%%%%%%%%%%%%%%%%%%
\input epsf.tex
\centerline{\epsfbox{fig1.eps}}
\centerline{{\bf Figure: }
Surviving Generators}
%%%%%%%%%%%%%%%%%%%%%%%%

%%%%%%%%%%%%%%%%%%%%%%%%%%%%%%%%%%%%%%%%%%%%%%%%%%%
\subsection{Definition of the \glinfh algebra}
In order to prevent the appearance of infinity
once we try to relate it with \Winf, we need to
modify the generators of \glinf as follows:
\be
E^{(\mu)}(i,j) = \Eb^{(\mu)}(i,j)-
c^{(\mu)}\delta_{i+j,0}\theta(i).
\label{hatgen}
\ee
Here the ``central charges'' are defined by,
\be
c^{(\mu)}=\sum_{k\in S^{(\mu)}} h^{(\mu)}_k.
\label{ccs}
\ee
The algebra \eq{glinf} and the
highest weight condition \eq{gl-hwc}
are also modified,
\ba
\left[ E^{(\mu)}(i,j), E^{(\nu)}(k,\ell)\right]
&=&\theta(m-\mu-\nu)\times\n
&&\left(\delta_{j+k,0}E^{(\mu+\nu)}(i,\ell)-
\delta_{i+\ell,0}E^{(\mu+\nu)}(k,j)
\right.\n
&&\left. +c^{(\mu+\nu)}\delta_{j+k,0}\delta_{\ell+i,0}
(\theta(i)-\theta(k))\right),
\label{glcom}
\ea
and,
$E^{(\mu)}(i,-i)\vac = q^{(\mu)}_i \vac$
with $q^{(\mu)}_i= \qb^{(\mu)}_i -c^{(\mu)}\theta(i)$.
We can also easily prove,
$$h_k^{(\mu)}= q^{(\mu)}_k
-q^{(\mu)}_{k-1}+ c^{(\mu)}\delta_{k,0}$$
We call the modified algebra \eq{glcom} as the \glinfh algebra.
We remark that the quasifinite representations of \glinf
and \glinfh are identical since there appear no infinite sum
in the definition.

%%%%%%%%%%%%%%%%%%%%%%%%%%%%%%%%%%%%%%%%%%%%%%%%%%%%%%%%%%%%%%%%%%%
\subsection{Relation with the \Winf algebra}
To find a relation between \glinfh and \Winf, we take the Hilbert
space spanned by $\vv_k$ as the space of functions on the circle
spanned by $z^{\lambda+k+t}$ with $\lambda \in {\bf C}$, $k\in \ZZ$.
Here the formal parameter $t$ is defined by
nilpotency condition, $t^{m+1}\equiv 0$.
The action of differential operators on this basis is then given by,
\ba
z^r f(D) z^{\lambda +k+t} &=&
f(\lambda +k+t) z^{\lambda+k+r+t}\n
&=&\left( \sum_{\mu=0}^m \frac{t^{\mu}}{\mu!}
f^{(\mu)}(\lambda+k)\right) z^{\lambda+k+r+t}.
\label{waction}
\ea
By the identification, $\vv^{(\mu)}_k \leftrightarrow
t^{\mu}z^{\lambda+k+t}$, we define the correspondence between
the generators,
\be
W(z^r f(D))= \sum_{k\in \ZZ}\sum_{\mu=0}^m
\frac{f^{(\mu)}(\lambda +k )}{\mu !}
E^{(\mu)}(r+k,-k),
\label{3.2}
\ee
for $r\neq 0$.
Special care is needed to find the relation between the zero modes,
\ba
W(e^{xD})&=& \sum_{k\in \ZZ}\sum_{\mu=0}^m
\frac{x^\mu e^{xk}}{\mu!}\left(e^{\lambda x}\Eb^{(\mu)}(k,-k)
-\delta_{\mu,0}c^{(0)}\theta(k)\right)\n
&=&  \sum_{k\in \ZZ}\sum_{\mu=0}^m
\frac{x^\mu e^{xk}}{\mu!}e^{\lambda x}E^{(\mu)}(k,-k)\n
&&-c^{(0)} \frac{e^{\lambda x}-1}{e^x -1}
-\sum_{\mu=1}^m\frac{ x^{\mu} c^{(\mu)}e^{\lambda x}}{\mu! (e^x-1)}.
\label{3.4}
\ea
The central charges of the both algebras ($C$ for \Winf and
$c^{(0)}$ for \glinfh) are related by,
\be
C=c^{(0)}.
\label{cc}
\ee
The other central charges of \glinfh, $c^{(\mu)}$,
can be related to the coefficients of $x^\mu$ in the polynomial
$p_{K_i}(x)$ in \eq{eDelta} as we see in the next subsection.

%%%%%%%%%%%%%%%%%%%%%%%%%%%%%%%%%%%%%%%%%%%%%%%%%%%%%%%%%%%%%%%%%
\subsection{Relation between the representations}
In the correspondence \eq{3.2}, $\lambda$
is a free parameter.  This arbitrariness is removed once
we consider the relation between the (quasifinite)
representations of \Winf and \glinfh.

Let us examine the null state conditions,
$W(z^{-r}b_r(D))\vac =0$ in the language of \glinfh.
We first consider the case,
\be
b(w)=\prod_{i=1}^\kappa (w-\lambda'-k_i)^{\mu_i},
\label{spb}
\ee
i.e.
the differences of
all the root of characteristic polynomial
are integers.
We put $m'\equiv \mbox{max}(\mu_i)$.
We introduce the set of integers associated with $b$ as,
$$
T^{(\mu)}\equiv \left\{
k\in \ZZ| b^{(\mu)}(\lambda' +k)=0\right\}.
$$
It is obvious that they are determined uniquely from $b(w)$
and satisfy the inclusion relation,
$$T^{(0)}\supseteq T^{(1)} \supseteq \cdots \supseteq T^{(m')},$$
which is the same as \eq{incl}.

Let $\lambda$ in \eq{3.2} be equal to $\lambda'$
and $m'=m$,
the null state condition,
\be
0=W(z^{-r}b_r(D))\vac=
\sum_{k\in \ZZ}\sum_{\mu=0}^m \frac{b_r^{(\mu)}(\lambda +k)}{\mu!}
E^{(\mu)}(-r+k,-k)\vac
\ee
implies that only the states of the form,
$$
E^{(\mu)}(k-1,-k)\vac
$$
with $k\in T^{(\mu)}$ may have non-vanishing norm
since the coefficient in \eq{3.2} vanishes.
This condition is identical with the quasifinite representation
of \glinf with,
\be
S^{(\mu)}=T^{(\mu)}.
\label{crspd}
\ee

%The check of null state for higher level is similar and
%it reproduces the same Hilbert space if $b_r(w)$
%is given by \eq{I6}.

Another check to see the direct relation
between the representations of \Winf and \glinfh
is to calculate the function $\Delta(x)$ from the
highest weight of \glinfh.
We observe that the weight $q^{(\mu)}_k$
is given by,
$$q^{(\mu)}_k= \sum_{s\in S^{(\mu)}}
h^{(\mu)}_s \theta(k-s)-c^{(\mu)}\theta(k).$$
Combine it with \eq{3.4} and $\Delta(x)\vac=-W(e^{xD})\vac$,
we derive,
\be
\Delta(x)= \sum_{k\in S^{(0)}}
\frac{h^{(0)}_k(e^{x(\lambda+k)}-1)}{e^x-1}
+ \sum_{\mu=1}^{m}\sum_{k \in S^{(\mu)}}
\frac{x^{\mu}h^{(\mu)}_ke^{x(\lambda+k)}}{\mu! (e^x-1)}.
\label{detx}
\ee
This is exactly the solution of the differential equation
\eq{eDE} where $T^{(\mu)}$ of $b(w)$ is given by $S^{(\mu)}$.

In this way, the quasifinite representation of
\glinf we have seen in section 3.1 can be identified with
the representation of \Winf with the characteristic polynomial
\eq{spb}.

%%%%%%%%%%%%%%%%%%%%%%%%%%%%%%%%%%%%%%%%%%%%%%%%%%%%%%%%%%%
\subsection{Characteristic polynomial
            for quasi-finite representation}
%%%%%%%%%%%%%%%%%%%%%%%%%%%%%%%%%%%%%%%%%%%%%%%%%%%%%%%%%%%%

By the correspondence with the \glinf module,
the characteristic polynomials at higher levels
for the quasifinite representation should be
uniquely determined\footnote{
We have to remark that
the definition of the characteristic polynomial $b_r(D)$
in this paper is slightly different from \cite{rKR}.
In \cite{rKR}, it is
introduced associated with the parabolic subalgebras.
In this context, there are some arbitrariness
from its consistency condition alone.
On the other hand, we define $b_r(D)$
as the minimal monic polynomial that satisfies
$W(z^{-r}b_r(D))|\lambda\rangle= \mbox{null}$
for the generic $C$s.
}.
We claim,
%%%%%%%%%%%%%%%%%%%%%%%%%%%%%%%%%%%%%%%%%%%
%           LCM Theorem
%%%%%%%%%%%%%%%%%%%%%%%%%%%%%%%%%%%%%%%%%%%
\begin{proposition}{{\bf Irreducible qusifinite module}}

\noindent For the generic values of $C_i$,
if the weight function $\Delta(x)$ satisfies $\eq{eDE}$,
then there exist a unique irreducible quasifinite module such that
the characteristic polynomial is
\be
b_\nr(D)=\mbox{l.c.m.}(b(D), b(D-1),\cdots, b(D-\nr+1)),
\label{I6}
\ee
where $b(w)$ is a minimal degree
monic polynomial satisfing $\eq{eDE}$.
\end{proposition}
%==================================================

\noindent{\bf Proof:}
This follows from the following lemma and
the relation with the $gl(\infty)$ given in this section.
\qed

%%%%%%%%%%%%%%%%%%%%%%%%%%%%%%%%%%%%%%%%%%%
%           Null state lemma
%%%%%%%%%%%%%%%%%%%%%%%%%%%%%%%%%%%%%%%%%%%
\begin{lemma}{{\bf Null state}}

\noindent
If the weight function $\Delta(x)$ satisfies $\eq{eDE}$,
then the state
$W(z^{-\nr}e^{yD}f(D))\KET$ with
\be
f(D)=\mbox{l.c.m.}(b(D), b(D-1),\cdots, b(D-\nr+1))
%\label{I6}
\ee
is a null state.
\end{lemma}
%==================================================

\noindent{\bf Proof:}
By combining relations,
$$W(z^{-\nr} e^{yD}f(D))=f\left(\sfrac{d}{dy}\right)W(z^{-\nr}e^{yD})$$ and
\ba
 && \left[ W(z^\ms e^{x(D+\ms)}), W(z^{-\nr} e^{yD})\right]\n
 &&~\qquad\qquad =(e^{x(\ms-\nr)}-e^{(x+y)\ms})
\left\{ W(z^{\ms-\nr}e^{(x+y)D})+\frac{C\delta_{\ms,\nr}}{1-e^{x+y}}\right\},
\ea
we can derive the following equation
for all $\ms_i$, $\nr\in\bZ_{>0}$ with  $\sum_{i=1}^{\rn} \ms_i=\nr$:
\ba
\BRA W(z^{\ms_\rn}e^{x_\rn(D+\ms_\rn)})\cdots W(z^{\ms_1}e^{x_1(D+\ms_1)})
W(z^{-\nr}e^{yD}f(D))\KET\qquad\qquad\quad~&&\n
=f\left(\sfrac{d}{dy}\right)\prod_{j=1}^{\rn}
\left( e^{x_j(\ms_{1,j}-\nr)} - e^{(x_{1,j}+y)\ms_j}\right)
%\n &&\times
\left\{-\Delta(x_{1,\rn}+y) + \frac{C}{1-e^{(x_{1,\rn}+y)}}\right\},
\label{nulli}
\ea
where $\ms_{1,j}\equiv\sum_{i=1}^j \ms_i$ and $x_{1,j}\equiv\sum_{i=1}^j x_i$.
If we set $X=x_{1,\rn}+y$,
then the right hand side of eq.\ \eq{nulli} reduces to
\be
\sum_{m=0}^{r-1}a_m(x,s)\,\,f\left(\sfrac{d}{dX}\right)
e^{mX}(1-e^X)\left\{-\Delta(X) + \frac{C}{1-e^{X}}\right\},
\label{nullll}
\ee
with some functions $a_m(x,s)=a_m(x_1,\cdots,x_n,s_1,\cdots,s_n)$.
On the other hand, the differential equation \eq{eDE}
may be rewritten as
\be
b(\sfrac{d}{dx}-\ms)e^{\ms x}
\left(1-e^{x}\right)\left(-\Delta(x)+\frac{C}{1-e^x}\right)=0.
\ee
%If we define a monic polynomial $d_\nr(D)$ as,
%$$d_\nr(D)\equiv\mbox{l.c.m.}(b(D), b(D-1),\cdots, b(D-\nr+1)),$$
%it satisfies\be d_\nr(\sfrac{d}{dx})e^{\ms x}
%\left(1-e^{x}\right)\left(-\Delta(x)+\frac{C}{1-e^x}\right)=0,\ee
%for $\ms=0,1,\cdots,\nr-1$. If we replace $f(D)$ by $d_\nr(D)$,
Hence, the right hand side of eq.\ (\ref{nulli}) vanishes.
Since any inner product with
$W(z^{-\nr}e^{yD}f(D))\KET$ is zero,
it is a null state.
%By definition, it means that $b_\nr(D)$ divides $d_\nr(D)$.
%However, as Kac-Radul observed, $d_\nr(D)$ also divides $b_\nr(D)$.
%It means that they are identical.
\qed

%%%%%%%%%%%%%%%%%%%%%%%%%%%%%%%%%%%%%%%%%%%%%%%%%%%%%%%%%%%%%%%
%
%   Section 4 \lambda dependence
%   This section is NEW!!
%
%%%%%%%%%%%%%%%%%%%%%%%%%%%%%%%%%%%%%%%%%%%%%%%%%%%%%%%%%%%%%%%

\section{$\lambda$ dependence}

We divide the derivation of the determinant formula \eq{2.2}
into two parts. As we have reviewed in the previous section,
at each level, the same is the dimensions of
the generalized Verma module of \Winf and \glinf algebras.
Let us consider the Hilbert space at a spcific energy level.
We denote $\{ u_1, \cdots, u_N\}$ as the basis
in terms of \Winf generators and
$\{ v_1, \cdots v_N\}$ as the basis
in terms of \glinf generators.
The relation between the two basis may be written as,
$u_i = \sum_j {\cal A}_{ij} v_j$, with $N\times N$ matrix
${\cal A}$.
The matrix ${\cal A}$ can be directly derived from the relation
\eq{3.2} and it depends only on $\lambda_i$s.
The determinant for the \Winf basis is rewritten
as the determinant for the \glinf generators.
\be
	Det(\langle u_i| u_j \rangle)=
	Det({\cal A})^2 Det(\langle v_i| v_j \rangle)
	\label{L1}
\ee
In the representation of the \glinf algebra, the
only parameters which appear in the theory are
% $h^{(\mu)}_k$ which are what we wrote as
$C_i$s.  This observation shows that \eq{L1} gives a natural
decomposition of the determinant into a part
which depends only on $\lambda_i$s ($Det({\cal A})^2$),
and a part which depends only on $C_i$s
($Det(\langle v_i| v_j \rangle)$).
In this section, we derive the first factor.

The main theorem in this section is
%%%%%%%%%%%%%%%%%%%%%%%%%%%%%%%%%%%%%%%%%%%%%%%%%%%%%%%%%%%
\begin{theorem}{\bf $\lambda$ dependence}
\label{th-lambda}
\newline
\noindent
The factor $B_r$ in \eq{2.2} is given by,
\be
B_r(\lambda) =\left(
\lambda^{\mu_0^{(r)}} \prod_{s=1}^\infty
(\lambda+s)^{\mu_s^{(r)}}(\lambda-s)^{\mu_s^{(r)}}
\right)^2.
\label{L2}
\ee
Here the non-negative integers $\mu_s^{(r)}$ can be
derived from the generating function,
\be
\phi_s(q)\equiv
\sum_{r=0}^\infty \mu_s^{(r)} q^r
=
\left.\frac{\partial}{\partial \zeta}
\left( \left( \prod_{r=1}^\infty \frac{1}{(1-q^r)^r}\right)^K
\prod_{t=s+1}^\infty \frac{(1-q^t)^{t-s}}{(1-\zeta
q^t)^{t-s}}\right)\right|_{\zeta=1}.
\label{L3}
\ee
\end{theorem}
%%%%%%%%%%%%%%%%%%%%%%%%%%%%%%%%%%%%%%%%%%%%%%%%%%%%%%%%%%%
\vskip 5mm
{}For the simplest cases, $K=2$
with
$\lambda_1-\lambda_2= 0,\pm 1$,
\eq{L3} gives respectively,
\ba
2 \phi_0(q)
&=&
2q+10q^2+34q^3+108 q^4+298 q^5\cdots,\n
2 \phi_1(q)
&=&2q^2+8q^3+30 q^4+88 q^5+\cdots,\nonumber
\ea
which correctly reproduce the table in appendix A.

Before we start the detailed proof, it may be
useful in the future study to give the intuitive proof
of this theorem.

{}From \eq{I6}, the factor
$t-s$ in \eq{L3} can be regarded as
the number of additional null generators at level $t$
when a pair $(\lambda_i, \lambda_j)$
satisfies the relation
$\lambda_i-\lambda_j=\pm s$.
More explicitly,
\be
t-s=tK-\mbox{degree }
(\mbox{l.c.m}(b(w),b(w-1),b(w-2),\ldots,b(w-t+1))).
\label{l5}
\ee
If a state in the Verma module has the form,
$N(\bullet)^m \cdot W(\bullet)^n\vac$,
where $N(\bullet)$s are any null operators,
the inner product of this state with any bra state will
 get a factor $(\lambda_i-\lambda_j-s)^m$.
In order to collect the power factor $m$ for the all
state at energy level $t$,  we attach a factor $\zeta$ with $q$
in order to mark the null generators.
A state of the form, $N(\bullet)^m \cdot W(\bullet)^n\vac$,
will get a factor $\zeta^m$.  We take a derivative
with respect to $\zeta$ to pick up the multiplicity factor $m$.
Since the bra states should get the same factor,
we multiply the coefficient of $q^m$ by two.
This argument shows that the determinant
can be divisible by the factor in \eq{L2}.

\subsection{Relation betweem generators}
The proof of the theorem is straightforward but
a little lengthy. We will divide the argument into  small steps.

The independent generators
in  the \Winf algebra at level $r$ are $W(z^{-r} D^s)$
with $s=0,1,\cdots, rK-1$. On the other hand, those in the \glinf
algebra may be taken as $E_{\lambda_s}(-r+j, -j)$
with $s=1,\cdots, K$ and $j=0,\cdots, r-1$.
We denote the \glinf generator associated
with parameter $\lambda$ as $E_{\lambda}(i,j)$.
{}From \eq{3.2}, those generators are related by $rK\times rK$ matrix $A_r$
as,
$W(z^{-r}D^i)=\sum_{s=1}^K \sum_{j=0}^{r-1}
(A_r)_{i,(s-1)r+j}E_{\lambda_s}(-r+j, -j)$ with the matrix element,
\be
 (A_r)_{i,(s-1)r+j}=(\lambda_s+j)^i.
 \label{L4}
\ee
The  matrix $A_r$
has the form of the Vandermonde matrix. It is hence quite easy
to derive its determinant as,
%%%%%%%%%%%%%%%%%%%%%%%%%%%%%%%%%%%%%%%%%%%%%%
\begin{lemma}{}
Up to the multiplication of constant,
\ba
\mbox{det}(A_r)&=&
\prod_{1\leq i<j\leq K}\prod_{k,\ell=0}^{r-1}
(\lambda_i-\lambda_j+k-\ell)
\nonumber\\
&=&
\prod_{1\leq i<j\leq K}(\lambda_i-\lambda_j)^r \prod_{\epsilon=\pm1}
\prod_{s=1}^{r-1}(\lambda_i-\lambda_j+\epsilon s)^{r-s}.
\label{vandermonde}
\ea
\end{lemma}
%%%%%%%%%%%%%%%%%%%%%%%%%%%%%%%%%%%%%%%%%%%%%%%%%%%%

%%%%%%%%%%%%%%%%%%%%%%%%%%%%%%%%%%%%%%%%%%%%%%%%%%%%
\subsection{Relation between Hilbert spaces}
%%%%%%%%%%%%%%%%%%%%%%%%%%%%%%%%%%%%%%%%%%%%%%%%%%%%%

Let ${\cal H}(n_1, n_2,n_3, \cdots)$ be the Hilbert space
spanned by the product of $n_1$ elements of level 1 generators,
$n_2$ elements of level 2 generators, $n_3$ elements of level 3
generators and so on, which are acting on the highest weight state.
Here, $n_i$s are the non-negative integers.
In order to consider the determinant at finite level,
only finite number of them can be non-vanishing.
The energy level can be written out of them as,
\be
N=\sum_{\ell=1}^\infty \ell n_\ell.
\label{L5}
\ee

The basis of this Hilbert space
 may be written either in terms of the \Winf generators
 or in terms of the \glinf generators.
The transformation matrix between those basis can be constructed
out of the matrices $A_r$ which is introduced in the
previous subsection.
{}For the Hilbert space ${\cal H}(n_1, n_2, \cdots)$, it is
given by,
\be
A_1^{(n_1)}\otimes A_2^{(n_2)}\otimes A_3^{(n_3)} \otimes \cdots
=\bigotimes_{r=1}^\infty A_r^{(n_r)},
\label{L6}
\ee
where $A_r^{(n_r)}$ is the transformation matrix between
the $n_r$-th symmetrized
product of the original $rK$ basis.
In this case,  $A_r^{n_r}$ becomes
\newline
$\left( \begin{array}{c} rK+n_r-1\\ n_r \end{array}\right)
\times
\left( \begin{array}{c} rK+n_r-1\\ n_r \end{array}\right) $
matrix.

To derive the determinant for the matrix \eq{L6},
we need to remark some identities of the linear algebra,
which can be proved easily.
%%%%%%%%%%%%%%%%%%%%%%%%%%%%%%%%%%%%%%%%%%%%%%%%%

\begin{lemma}{ }
%\noindent
(1) Let $B_r$ be an arbitrary $N_r \times N_r$
matrix $(r=1,2,\cdots, M)$.
The determinant for the direct product matrix is given by,
\be
\mbox{det}(B_1\otimes\cdots \otimes B_M)=
\prod_{r=1}^M (\mbox{det}B_r)^{\nu_r},
\qquad
\nu_r=
{(\prod_{s=1}^M n_s)/n_r}.
\label{L7}
\ee

\noindent (2)
Let $B$ be an arbitrary $N \times N$ matrix.
If we denote $B^{(M)}$ as the representation of
$B$ in terms of $M$th symmetric basis. Then,
\be
\mbox{det} B^{(M)}=(\mbox{det} B
)^{\sigma_M},
\qquad
\sigma_{M}=
{\left( \begin{array}{c} N+M-1\\ N \end{array}\right)}.
\label{L8}
\ee

\end{lemma}
%%%%%%%%%%%%%%%%%%%%%%%%%%%%%%%%%%%%%%%%%%%%%%%%%%%%%%

The deter\-minant formula for the space
${\cal H}(n_1, n_2, n_3, \cdots)$ is
derived as,
%%%%%%%%%%%%%%%%%%%%%%%%%%%%%%%%%%%%%%%%%%%%%%%%%%%%%%
\begin{lemma}{}
%\newline
\be
\mbox{det}(\bigotimes_{r=1}^\infty
A_r^{(n_r)})
=\prod_{r=1}^\infty (\mbox{det}A_r)^{q_r([n])},
\label{L8.5}
\ee
with
$$q_r([n]) =\left( \begin{array}{c} n_r+rK-1\\ rK
\end{array}\right)\prod_{
{\scriptstyle s=1} \atop {\scriptstyle  s\neq r}
}^\infty
\left( \begin{array}{c} n_s+sK-1\\ n_s \end{array}\right).$$
If we use \eq{vandermonde}, this formula becomes,
\be
\prod_{i<j}\left( (\lambda_i - \lambda_j)^{\alpha_0([n])}
\prod_{s=1}^\infty
((\lambda_i-\lambda_j+s)(\lambda_i-\lambda_j+s))^{\alpha_s([n])}
\right)
\label{L9}
\ee
with
\be
\alpha_s([n])=\sum_{t=s+1}^\infty (t-s)
\left( \begin{array}{c} n_t+tK-1\\ tK \end{array}\right)
\prod_{
{\scriptstyle u=1} \atop {\scriptstyle  u\neq t}
}^\infty
\left( \begin{array}{c} n_u+uK-1\\ n_u \end{array}\right).
\label{L10}
\ee
\end{lemma}

%%%%%%%%%%%%%%%%%%%%%%%%%%%%%%%%%%%%%%%%%%%%%%%%%%%%%
\subsection{Generating function}
%%%%%%%%%%%%%%%%%%%%%%%%%%%%%%%%%%%%%%%%%%%%%%%%%%%%%

{}Finally, to derive the generating functional \eq{L3},
we take the summation over infinite indices $(n_1,
n_2,\cdots)$ with parameter $q$,
\be
\phi_s(q) =
\sum_{n_1, n_2, \cdots} \alpha_s([n])
q^{\sum_{j=1}^\infty j n_j}.
\label{L11}
\ee
Combining it with \eq{L10}, and by using the Taylor expansions,
\ba
\frac{1}{(1-q^u)^{Ku}}& = &
\sum_{n=0}^\infty
\left( \begin{array}{c} n+uK-1\\ n \end{array}\right) q^{un}
\nonumber\\
\frac{q^t}{(1-q^t)^{Kt+1}}& = &
\sum_{n=1}^\infty
\left( \begin{array}{c} n+tK-1\\ tK \end{array}\right) q^{tn},
\label{L12}
\ea
we get the explicit form of the summation,
\ba
\phi_s(q) & = &
\sum_{t=s+1}^\infty \frac{(t-s)q^t}{(1-q^t)^{Kt+1}}
\prod_{
{\scriptstyle u=1} \atop {\scriptstyle  u\neq t}
}^\infty \frac{1}{(1-q^u)^{Ku}}
\nonumber\\
& = &
\left.
\frac{\partial}{\partial \zeta}
\left( \left( \prod_{r=1}^\infty \frac{1}{(1-q^r)^r}\right)^K
\prod_{t=s+1}^\infty \frac{(1-q^t)^{t-s}}{(1-\zeta q^t)^{t-s}}\right)
\right|_{\zeta=1}.
\label{L13}
\ea
It completes our derivation of theorem \ref{th-lambda}.
\qed

%%%%%%%%%%%%%%%%%%%%%%%%%%%%%%%%%%%%%%%%%%%%%%%%%%%%%%%%%%%%%%
%
%               Section 4 C dependence
%
%%%%%%%%%%%%%%%%%%%%%%%%%%%%%%%%%%%%%%%%%%%%%%%%%%%%%%%%%%%%%%
\section{$C$ dependence}
In the following section, we derive the $C$ dependence of
the determinant formula for the case, $K=1$,
$b(w)=w-\lambda$, $\Delta(x)=C \frac{e^{\lambda x}-1}{e^x-1}$.
The computation is basically carried out by using the
\glinfh generators.  The relation between the nonvanishing
generators contain the dependence
on $\lambda$ but it will disappear
if we take the determinant. Hence the determinant formula
of \glinfh is identical with that of \Winf.

Computation for $K=1$ is sufficient for understanding
the result in our previous computation appendix A,
since  they are the direct product of $K=1$ representations.

Unfortunately, the determinant formula for
more non-trivial cases, where the characteristic polynomial
has roots whose mutual difference is an integer,
is still beyond the scope of the present paper.

%%%%%%%%%%%%%%%%%%%%%%%%%%%%%%%%%%%%%%%%%%%%%%%%%%%%%%%%%%%%%%
\subsection{Classification by the complete Cartan elements}
There are an infinite number of commuting charges
(forming the Cartan subalgebra) in the \Winf algebra,
$W(D^k)$ with $k=0,1,2,\ldots$.
In our previous computation, we used only $L_0 \equiv -W(D)$ to
classify states.  However, much more
detailed analysis should be possible
if we diagonalize the Hilbert state
with respect to the action of all the Cartan elements.

In the framework of the \Winf algebra, however,
the construction of the Weyl basis is not so straightforward
since
a simple commutation shows that
\be
\left[W(D^k), W(z^r f(D))\right]
= W(z^r ((D+r)^k-D^k)f(D)).
\label{3.10}
\ee
It is obvious that
we need to diagonalize the operator $Q$
which acts on the \Winf generator as
$$Q[W(z^r f(D))]=W(z^r Df(D)).$$
If we restrict $f(D)$ to be
a polynomial, we can not find any solution
to this equation.

The construction of the diagonal basis becomes possible
if we view the \Winf algebra from the equivalent \glinfh algebra.
In \cite{rKR}, they proved that the quasifinite representation
of those algebras coincide.

In the language of \glinfh,
the generators $E^{(0)}(i,j)$ are already
diagonal with respect to the action of the Cartan
elements\footnote{Here and in the following discussion, we
omit the superscript $(0)$ in $E(i,j)$ since we are
only considering $K=1$ cases.},
\be
\left[W(D^k),E(i,j)\right]
= ((\lambda+i)^k-(\lambda-j)^k)E(i,j).
\label{3.12}
\ee
The state $E(-i_1,-j_1)\cdots
E(-i_n, - j_n)\vac$ has the eigenvalue,
\be
\sum_{a=1}^n \,\left[\,(\lambda-i_a)^k
-(\lambda+j_a)^k\,\right]\,
+\,\Delta_k,
\label{EGV}
\ee
with respect to the action of $W(D^k)$.
To summarize, we may claim (for $K=1$ case),
%%%%%%%%%%%%%%%%%%%%%%%%%%%%%%%%%%%%%%%%%%%%%%%%%
\begin{proposition}{\bf Classification of states}
\newline
\noindent
Let $I\equiv \left\{i_1,\cdots,i_n\right\}$
$($resp. $J\equiv \left\{j_1,\cdots,j_n\right\}$$)$
be a set of positive $($resp. non-negative$)$ integers
and $\sigma$ be a permutation of the set of integers $1,\cdots,n$.
The eigenvectors with respect to all $W(D^k)$ are given as
the linear combinations of the form,
\be
\sum_\sigma c_\sigma\prod_{a=1}^n E(-i_a, -j_{\sigma(a)})\vac,
\label{basis}
\ee
with $c_\sigma\in\CC$.
\end{proposition}
%================================================
%\begin{theorem}{\bf Classification of states}
%\newline\noindent
%The subspace of the {\rm \glinfh}module with identical eigenvalues
%with respect to all $W(D^k)$ is spanned by the basis,
%\be\prod_{a=1}^n E(-i_a, -j_{\sigma(a)})\vac,\label{basis}\ee
%where $I\equiv \left\{i_1,\cdots,i_n\right\}$
%$($resp. $J\equiv \left\{j_1,\cdots,j_n\right\}$$)$
%is a set of positive $($resp. non-negative$)$ integers
%and $\sigma$ is a permutation of the set of integers
%%$1,\cdots,n$.\end{theorem}
%
%%%%%%%%%%%%%%%%%%%%%%%%%%%%%%%%%%%%%%%%%%%%%%%%%%%%%%%%%
\subsection{Explicit calculation of inner product}
Due to the above theorem, we understand
that we need to consider only the class of
states of the form \eq{basis} to diagonalize the
Hilbert space.  For that purpose, we would like to prove
the explicit form of inner product between those
states,
\be
\bvac \prod_{a=1}^n (E(j_{\sigma(a)},i_a)
\prod_{b=1}^n E(-i_b, -j_{\sigma'(b)}))\vac=
(-1)^n (-C)^{L(\sigma^{-1}\sigma')}.
\label{inpro}
\ee
This equation is valid if all the indices $i$ (or $j$) are
given by different integers.
The function $L(\sigma)$ is the
``depth" of the  permutation $\sigma$.
It is known that any element of the permutation group
can be written as the product of cycles. For example,
\be
\left(
\begin{array}{cccccc}
1&2&3&4&5&6\\
3&6&4&1&5&2
\end{array}
\right)
=
(134)(26)(5).
\label{d11}
\ee
The function $L(\sigma)$ is then given by the number of
the cycles (including trivial one cycle).
In the above example, $L(\sigma)=3$.

In order to prove \eq{inpro},
we observe that, due to the nature of the \glinfh algebra,
the indices which appear in \eq{inpro} may be replaced by
other integers,
$$
\bvac \prod_{a=1}^n E(a,a)
\prod_{b=1}^n E(-b, -\sigma^{-1}\sigma'(b))\vac.
$$
In the following, we will write $\sigma^{-1}\sigma'$ as $\sigma$
for simplicity.
Let us first consider the case  $L(\sigma)=1$.
We postulate that
the inner product that consists of
a cycle of length $m$ is given by
$(-1)^m (-C)$ up to $m=n-1$
and prove the statement by induction.
This assumption is straightforwardly
proved for $m=1$
since the only non-vanishing contribution
comes from the central charge of the algebra.
The typical element which consists of
one cycle with $n$ element may be taken as,
$$
\bvac E(n,n)\cdots E(1,1)E(-1,-2)E(-2,-3)
\cdots E(-n,-1)\vac .
$$
We move the element $E(1,1)$ to the right.
Non-vanishing commutation relation happens
only with $E(-1,-2)$ and $E(-n,-1)$,
generating $E(1,-2)$ and $-E(-n,1)$, respectively.
However, the latter one vanishes after it is operated
on the vacuum.
Next, we move thus obtained element $E(1,-2)$ to the right.
This time, only nontrivial element is the commutation with
$E(-n,-1)$.  It gives the contribution $-E(-n,-2)$.
In this way, one arrives at the expression,
$$
-\bvac E(n,n)\cdots E(2,2)E(-2,-3)
\cdots E(-n,-2)\vac .
$$
However, this is the inner product which consists
of one cycle with $n-1$ element.
By induction assumption, it is equal to
$(-1)^n (-C)$.

If there are several cycles, the argument
similar to the above can be used to reduce the inner
product to the product of cycles.  Therefore, we have,
$$
\prod_{cycles} (-1)^m (-C) = (-1)^n (-C)^{L(\sigma)}
%(-1)^{(n-L(\sigma))}
%\bvac E(1,1)\cdots E(L(\sigma),L(\sigma))
%E(-1,-1)\cdots E(-L(\sigma),-L(\sigma))\vac.
$$
%\qed

%%%%%%%%%%%%%%%%%%%%%%%%%%%%%%%%%%%%%%%%%%%%%%%%%%%%%%%%%%%%%%%%%%%%%%%

\newcommand{\Sn}{{{\cal S}_n}}

%%%%%%%%%%%%%%%%%%%%%%%%%%%%%%%%%%%%%%%%%%%%%%%%%%%%%%%%%%%%%%%%%%%%%%%
\subsection{Young Diagram Classification}

Since the inner product formula is written
in terms of the permutation group and its
representation, we can easily believe that
the diagonal basis is explicitly constructed
by organizing them such as to give the irreducible
representation of the permutation group.
To accomplish this, we first prepare some notations.

Let $\Sn$ be the permutation group for $n$ objects.
Conjugacy classes of $\Sn$ are classified according to the type of
cycle decomposition (as in eq.\ (\ref{d11})).
Denoting by $k_j$ the number of length-$j$ cycles,
we represent a conjugacy class
as $(k)=1^{k_1}2^{k_2}\cdots n^{k_n}$.
%The depth of the class $(k)$ is defined to be the number of
%nonvanishing $k_j$'s and denoted by $L(k)$.
Note that $k_1 + 2k_2 + \cdots + nk_n = n$, and
the number of elements in the class $(k)$
is $N(k)\equiv n! / (1^{k_1}k_1! 2^{k_2}k_2! \cdots n^{k_n}k_n!)$.
Irreducible representations are classified by Young diagrams $Y$,
and we denote the character and dimension of irreducible
representation $Y$ by $\chi_Y$ and $d_Y$, respectively.

We define the action of $\sigma \in \Sn$ on the state
$\prod_{a=1}^{n}E(-i_a,-j_a)\vac$ by
\be
   \sigma \prod_{a=1}^nE(-i_a,-j_a)\vac
   \equiv \prod_{a=1}^nE(-i_a,-j_{\sigma(a)})\vac.
\ee
We then introduce the operator
\be
   B^Y_{\alpha\beta} \equiv \frac{d_Y}{n!} \sum_{\sigma\in\Sn}
   D^Y(\sigma)_{\alpha\beta}\sigma.
\ee
Here $D^Y(\sigma)_{\alpha\beta}~(\alpha,\,
\beta=1,2,...,d_Y)$ is the
(real-valued) representation matrix of element $\sigma$.
Since $\sigma^\dagger=\sigma^{-1}$,
we obtain the following relations
(see, for example, \cite{me} for the proof):
\ba
   {B^Y_{\alpha\beta}}^\dagger&=&B^Y_{\beta\alpha}, \label{B1}\\
   B^Y_{\alpha\beta}B^{Y'}_{\mu\nu}&=&
   \delta_{YY'}\delta_{\beta\mu}B^Y_{\alpha\nu}. \label{B2}
\ea
With this operator we define new vectors as follows:
\ba
   \ket{Y;\alpha\beta}\equiv B^Y_{\alpha\beta}\prod_{a=1}^n
   E(-i_a,-j_a)\vac.
\ea
In the following, we will restrict our discussion to the case
where no degeneracy exists in the
set of indices, $\{j_a\}$ (and also
in $\{i_a\}$).

We are now in a position to prove the following theorem:

%%%%%%%%%%%%%%%%%%%%%%%%%%%%%%%%%%%%%%%%%%%%%%%%%%%%%%
\begin{theorem}{{\bf Young Diagram Classification}}
\newline
\noindent
The vectors $\ket{Y;\alpha\beta}$ form an orthogonal basis in the
subspace spanned by $\eq{basis}$:
\be
   \langle Y;\alpha\beta | Y';\mu\nu\rangle
   =\delta_{YY'}\delta_{\alpha\mu}\delta_{\beta\nu}
   a_n^{Y},
   \quad a_n^Y=
   \frac{d_Y}{n!}\prod_{b\in Y}(C-C_b). \label{f1}
\ee
Here to each box $b$ in the Young diagram,
we assign a number $C_b$ as,
\be
\begin{tabular}{|c|c|c|c|c}\hline
$0$&$1$&$2$&$3$&$\cdots$\\ \hline
$-1$&$0$&$1$&$2$&$\cdots$\\ \hline
$-2$&$-1$&$0$&$1$&$\cdots$\\ \hline
$-3$&$-2$&$-1$&$0$&$\cdots$\\ \hline
$\vdots$&$\vdots$&$\vdots$&$\vdots$&$\ddots$
\end{tabular}
\label{I11}
\ee
\end{theorem}
%===========================================

\noindent{\bf Proof:}
By using eqs.\ (\ref{B1}) and (\ref{B2}), the left-hand side of
eq.\ (\ref{f1}) is rewritten as
\ba
   &&\langle Y;\alpha\beta | Y';\mu\nu\rangle \nonumber\\
   &&~~~=\delta_{YY'}\delta_{\alpha\mu}\frac{d_Y}{n!}
         \sum_{\sigma\in\Sn}D^Y(\sigma)_{\beta\nu}
         \bra{\lambda}\prod_aE(j_a,i_a)\sigma
         \prod_bE(-i_b,-j_b)\vac.
\ea
Due to eq.\ (\ref{inpro}), $\bra{\lambda}\prod_aE(j_a,i_a)\sigma
\prod_bE(-i_b,-j_b)\vac=(-1)^n(-C)^{L(\sigma)}$.
Since $L(\sigma)$ is a class function, we may denote it by
$L(k)$ if $\sigma \in (k)$.
We thus obtain
\be
   \langle Y;\alpha\beta | Y';\mu\nu\rangle
   =\delta_{YY'}\delta_{\alpha\mu}\frac{d_Y}{n!}(-1)^n
    \sum_{(k)}(-C)^{L(k)}\sum_{\sigma\in(k)}D^Y(\sigma)_{\beta\nu}.
   \label{f2}
\ee
Here we can show that the matrix $\sum_{\sigma\in(k)}D^Y(\sigma)$
always commutes with the actions of any elements in $\Sn$,
and thus, due to Schur's lemma, we conclude that
$\sum_{\sigma\in(k)}D^Y(\sigma)$ is proportional to the unit matrix.
The coefficient is easily calculated by taking its trace, and we
obtain
\be
   \sum_{\sigma\in(k)}D^Y(\sigma)_{\beta\nu}
   =\frac{N(k)}{d_Y}\chi_Y(k)\delta_{\beta\nu}.
\ee
Substituting this expression into eq.\ (\ref{f2}), we obtain
\be
   \langle Y;\alpha\beta | Y';\mu\nu\rangle
   =\delta_{YY'}\delta_{\alpha\mu}\delta_{\beta\nu}
    \frac{(-1)^n}{n!}\sum_{(k)}(-C)^{L(k)}N(k)\chi_Y(k).
   \label{f3}
\ee
Since we have the following identity as is proved in appendix B.3:
\be
   \frac{(-1)^n}{n!}\sum_{(k)}(-C)^{L(k)}N(k)\chi_Y(k)
   =\frac{d_Y}{n!}\prod_{b\in Y}(C-C_b),
\ee
we finally obtain eq.\ (\ref{f1}).
\qed

As a simple corollary of the inner product formula,
we may derive the condition for the unitarity.
The positivity of the Hilbert space may be rephrased as
the positivity of the factor $a_n^Y$ for any $Y$ and $n$.
{}From the table \eq{I11}, we can immediately prove that
this condition is achieved only when $C$ is positive integer.

%%%%%%%%%%%%%%%%%%%%%%%%%%%%%%%%%%%%%%%%%
%
%        Section 5
%
%%%%%%%%%%%%%%%%%%%%%%%%%%%%%%%%%%%%%%%%%
\section{Character formulae for $K=1$ module}
In the previous section, we get the explicit form of the
norm of diagonal basis in terms of the Young diagrams.
To understand the structure of the Hilbert space,
we need to count the number of the states which belong
to each diagram and have the same eigenvalues
for all the Cartan elements.

The generating functional for such degeneracy is
neatly expressed by introducing the full character,
\be
\chi([g])\equiv \mbox{Tr}_{\cal H}
\exp\left(\sum_{k=0}^\infty g_k W(D^k)\right).
\label{I7}
\ee

{}For the $K=1$ module, if there are no null states
aside from those coming from characteristic polynomial,
the non-vanishing generators are given by
$E(-r,-s)$ with  $r\geq 1, s\geq 0$.
If we combine it with \eq{3.12}, we get the following theorem
%%%%%%%%%%%%%%%%% Theorem  %%%%%%%%%%%%%%%%%%%%%%%%%%%%%%%%%%%%
\begin{theorem}{\bf Full character for the generalized Verma module}

The full character for the generalized Verma module is
\ba
\chi([g])&=&
e^{\sum_{k=0}^\infty g_k \Delta_k}
\prod_{r=1}^\infty \prod_{s=0}^\infty \frac{1}{1-u_r v_s},
\label{I8}\\
&=& e^{\sum_{k=0}^\infty g_k \Delta_k}
\sum_{Y} \tau_Y(x)\tau_Y(y),
\label{I12}
\ea
where
\be
u_r\equiv
e^{\sum_{k=0}^\infty g_k(\lambda-r)^k},\qquad
v_s\equiv
e^{-\sum_{k=0}^\infty g_k(\lambda+s)^k},
\label{I9}
\ee
$\tau_{Y}$ is the character
of irreducible representation $Y$ of ${\rm gl}(\infty)$,
$($see appendix B.1$)$,
and the parameters $x$ and $y$ are
the Miwa variables for $u$ and $v$,
respectively:
\be
x_\ell={1\over\ell}\sum_{r=1}^\infty u_r^\ell,
\qquad
y_\ell={1\over\ell}\sum_{s=0}^\infty v_s^\ell, \qquad
\ell=1,2,3,\cdots.
\label{I13}
\ee
\end{theorem}
%==========================================================
$\Delta_k$ is defined in \eq{I3} with
$\Delta(x)=C\frac{e^{\lambda x}-1}{e^x-1}$.
The proof of \eq{I12} is given in appendix B.2.

If we expand \eq{I8} in (infinitely many) parameters $u_r$ and $v_s$ as
$$
\sum_{n=0}^\infty
\sum_{I_n,J_n}N(I_n,J_n)\prod_{i\in I_n}u_i\prod_{j\in J_n}v_{j},
$$
then $N(I_n,J_n)$ gives the number of the states of the form
\eq{basis}.
If we expand each factor in the summation of \eq{I12},
we can get the degeneracy with respect to each Young diagram $Y$,
and the eigenvalues.

{}For example, some of the simpler Schur polynomials
are expanded as follows:
\ba
\tau_{2}(x) & = & \frac{x_1^2}{2} +x_2=
\sum_{i<j} u_i u_j +\sum_{i} u_i^2,\n
\tau_{11}(x)&=& \frac{x_1^2}{2}-x_2
= \sum_{i<j} u_i u_j,\n
\tau_{3}(x)&=& \frac{x_1^3}{6}+ x_1 x_2 +x_3=
\sum_{i<j<k}u_iu_j u_k +\sum_{i\neq j}u_i^2u_j+\sum_iu_i^3,\n
\tau_{21}(x)&=&\frac{x_1^3}{3}-x_3=
2\sum_{i<j<k}u_i u_j u_k +\sum_{i\neq j} u_i^2 u_j,\n
\tau_{111}(x) &=&  \frac{x_1^3}{6}- x_1 x_2 +x_3=
\sum_{i<j<k}u_iu_j u_k.
\label{shur}
\ea

The result in the previous section shows that
the generalized Verma module for $K=1$
becomes reducible when
$C=$ integer.  We call this represenatation as
the {\em degenerate representation}.
The full character formula for the irreducible module
can be obtained by
combining previous theorem with \eq{f1}.
%%%%%%%%%%%%%%%%%%%%%% corollary %%%%%%%%%%%%%%%%%%%%%%%%%%%%%%
\begin{theorem}{{\bf Full Character of Degenerate Representations}}

\noindent
Let $V_n$ $($resp. $H_n )$ be the set of the Young diagrams
the number of whose columns $($resp. rows$)$ does not exceed $n$,
then the full characters of $C=\pm n$
% degenerate representations
are given by,
\ba
\chi_{C=n}&=&e^{\sum_{k=0}^\infty g_k \Delta_k}
\sum_{Y\in V_n}
\tau_Y(x)\tau_Y(y),\n
\chi_{C=-n}&=&e^{\sum_{k=0}^\infty g_k \Delta_k}
\sum_{Y\in H_n}
\tau_Y(x)\tau_Y(y).
\label{I14}
\ea
\end{theorem}
%==================================================

In the character formula for non-integer $C$ \eq{I12},
the summation is over every Young diagram, or in other words,
the two-dimensional sum.  On the other hand,
the character formula for degenerate representation \eq{I14},
the summation is restricted to one-dimensional indices.
The degeneracy of the Hilbert space
reduces the dimensionality of the system from three to two,
which naturally explains the hybrid nature of \Winf symmetry.

To make our formula \eq{I14}
into more familiar form, we give the explicit form of the
characters
which depend only on the parameter
$q$ associated with the eigenvalue of $L_0\equiv -W(D)$.
{}For this restriction, we replace
\be
u_r=q^r,
\qquad
v_r = q^r.
\ee
Namely, we put $g_k=-2\pi i\tau \delta_{k,1}$ with
$q=e^{2\pi i\tau}$.
The Miwa variables \eq{I13} are then rewritten as,
\be
x_\ell=\frac{1}{\ell} \frac{q^\ell}{1-q^\ell},
\qquad
y_\ell=\frac{1}{\ell} \frac{1}{1-q^\ell}.
\ee
After these replacements, a compact form of
the Schur polynomial can be given.
We introduce,
\ba
&&f_k(q;m_1,\cdots,m_k)\n
&&\equiv
\prod_{j=1}^{m_k}
	\frac{1}{(1-q^j)(1-q^{m_{k-1}+1+j})
	\cdots (1-q^{m_{k-1}+\cdots+m_1+k-1+j})}\n
&&=\prod_{j=1}^{m_k}\prod_{s=0}^{k-1}
\left(1-q^{\sum_{t=1}^s m_{k-t}+s+j}\right)^{-1}.
\ea
for non-negative integers $m_i$. When $m_k=0$,
we put $f_k(q;m_1,\cdots,0)=1$.
The Schur polynomial is then rewritten as,
\ba
\tau_Y (x)
&=&
q^{\sum_{j=1}^{n} \frac{j(j+1)}{2} m_j}
\prod_{k=1}^n
f_k(q;m_1,\cdots,m_k)\n
\tau_Y (y)
&=&
q^{\sum_{j=1}^{n} \frac{j(j-1)}{2} m_j}
\prod_{k=1}^n
f_k(q;m_1,\cdots,m_k)\n
\tau_{Y'} (x)
&=&
q^{\frac{1}{2}\sum_{j=1}^{n}
(\sum_{s=j}^n m_s)(\sum_{s=j}^n m_s+1)}
\prod_{k=1}^n
f_k(q;m_1,\cdots,m_k)\n
\tau_{Y'} (y)
&=&
q^{\frac{1}{2}\sum_{j=1}^{n}
(\sum_{s=j}^n m_s)(\sum_{s=j}^n m_s-1)}
\prod_{k=1}^n
f_k(q;m_1,\cdots,m_k).
\ea
where $Y=\left\{m_1+\cdots+m_n,
m_2+\cdots+m_n, \cdots , m_n\right\}$,
and $Y'$ is the transpose of the Young diagram $Y$.

The full character formulae \eq{I14} then give,
\ba
\chi_{C=n}(q)&=&q^{n\lambda(\lambda-1)/2}
\sum_{m_1=0}^\infty\cdots\sum_{m_n=0}^\infty
q^{\sum_{j=1}^{n} (\sum_{s=j}^n m_s)^2}\prod_{k=1}^n
f_k(q;m_1,\cdots,m_k)^2,\n
\chi_{C=-n}(q)&=&q^{-n\lambda(\lambda-1)/2}
\sum_{m_1=0}^\infty\cdots\sum_{m_n=0}^\infty
q^{\sum_{j=1}^{n} j^2 m_j}\prod_{k=1}^n
f_k(q;m_1,\cdots,m_k)^2,
\label{qchar}
\ea
with positive integer $n$.

In our previous letter  \cite{rAFMO1},
we gave the character formulae for $C=\pm 1$ and $C=n>0$.
{}For $C=\pm 1$ cases, our general formula \eq{qchar} gives,
\ba
\chi_{C=1}(q)&=&q^{\lambda(\lambda-1)/2}\sum_{m=0}^\infty
q^{m^2} \prod_{j=1}^m \frac{1}{(1-q^j)^2},\n
\chi_{C=-1}(q)&=&q^{-\lambda(\lambda-1)/2}\sum_{m=0}^\infty
q^{m} \prod_{j=1}^m \frac{1}{(1-q^j)^2},
\ea
which are exactly same as our previous formulae.
{}For $C=n>0$ cases, what we derived previously were,
\be
\chi_{C=n}(q)=q^{n\lambda(\lambda-1)/2}\prod_{j=1}^\infty
\prod_{k=1}^n (1-q^{j+k-1})^{-1}.
\ee
We have confirmed that it is equivalent to
\eq{qchar} by Taylor expansion
up to $q^{30}$.

%%%%%%%%%%%%%%%%%%%%%%%%%%%%%%%%%%%%%%%%%%%%%%%%%%%%%%%%%%
\section{Discussion}
Although we understand the representation of
\Winf to some extent, there are still many things
to be understood.
In particular, the $C$ dependence for $K>1$ is not still
well-understood. We hope to report on the full detail
in our  future issue.
In the mathematical side, we are currently working on
the supersymmetric extension \cite{rAFMO2},
the structure of the subalgebras \cite{rAFMO3}.
Those works will be related to the topological field theory
and/or  the matrix models.

The relation of the \Winf algebra with extended objects
seems also interesting from the geometrical viewpoint.
In this work, we used the basis $D^k$ to parametrize the
generators.  However, as Kac and Radul observed,
there is another parametrization of generators which
leads to different representation.  One example is
to use $q^{kD}$ basis.  One may regard it as the representation
based on torus (instead of sphere). In general, one
may imagine the possibility
of the representation theories based on higher-genus Riemann
surfaces.  If we want to apply the \Winf algebra to membranes,
for example, we need to consider the ``degenerate'' three
manifolds where such Riemann surface may appear.
The hybrid nature of the \Winf algebra that we have
observed in this paper may be important to
understand such phenomena.

After we submited this paper,
the full character for the unitary representation has
been given in \cite{rFKRW}.
One may check \cite{rAFMO5} that for the special case ($K=1$),
the formula obtained there coincides with ours
\eq{I14} with $C=n>0$.

%%%%%%%%%%%%%%%%%%%%%%%%%%%%%%
% definitions added by Odake %
%%%%%%%%%%%%%%%%%%%%%%%%%%%%%%
\newcommand{\f}{\flat}
\newcommand{\fb}{\bar{\flat}}
\newcommand{\dket}[1]{\|#1\rangle\!\rangle}
\newcommand{\dbra}[1]{\langle\!\langle#1\|}

%%%%%%%%%%%%%%%%%%%%%%%%%%%%%%%%%%%%%%%%%%%%
%%                                        %%
%%   Appendix A                           %%
%%                                        %%
%%%%%%%%%%%%%%%%%%%%%%%%%%%%%%%%%%%%%%%%%%%%
\section* {Appendix A: Determinant formulae at lower degrees}
\setcounter{section}{1}
\renewcommand{\thesection}{\Alph{section}}
\setcounter{equation}{0}

In this appendix, we give the explicit form of
the functions $A_\nr(C)$ and $B_\nr(\lambda)$
defined in \eq{2.2}
We can parametrize those functions in the form,
$$
  A_\nr(C)=\prod_{\ell \in \ZZ}(C-\ell)^{\alpha(\ell)},
  \qquad
  B_\nr(\lambda)=\prod_{\ell \in \ZZ}(\lambda-\ell)^{\beta(\ell)}
$$
We make tables for the index $\alpha(\ell)$ and $\beta(\ell)$.
We note that $\beta(\ell)=\beta(-\ell)$.  Hence we will write
them only for $\ell\geq 0$.

\noindent \underline{$K=1$}:\hskip 4mm $B_\nr\equiv 1$ due to the
spectral flow symmetry \cite{rAFMO1}.
\begin{center}
\begin{tabular}{|c||c|c|c|c|c|c|c|c|c|}\hline%\hline
  $\nr$&$\alpha(-1)$&$\alpha(0)$&$\alpha(1)$&$\alpha(2)$
  &$\alpha(3)$&$\alpha(4)$&$\alpha(5)$&$\alpha(6)$&$\alpha(7)$\\
  \hline
  1&0&1&0&0&0&0&0&0&0\\
  2&0&3&1&0&0&0&0&0&0\\
  3&0&6&3&1&0&0&0&0&0\\
  4&1&13&8&3&1&0&0&0&0\\
  5&3&24&17&8&3&1&0&0&0\\
  6&10&48&37&19&8&3&1&0&0\\
  7&23&86&71&41&19&8&3&1&0\\
  8&54&161&138&85&43&19&8&3&1\\
  \hline
\end{tabular}
\end{center}
%\vskip 10mm

\noindent \underline{$K=2$}
\begin{center}
\begin{tabular}{|c||c|c|c|c|c||c|c|c|c|}\hline%\hline
  $\nr$&$\alpha(-1)$&$\alpha(0)$&$\alpha(1)$&$\alpha(2)$
  &$\alpha(3)$&$\beta(0)$&$\beta(1)$&$\beta(2)$&$\beta(3)$\\
  \hline
  1&0&1&0&0&0&2&0&0&0\\
  2&0&4&1&0&0&10&2&0&0\\
  3&0&12&4&1&0&34&8&2&0\\
  4&1&34&14&4&1&108&30&8&2\\
  \hline
\end{tabular}
\end{center}
%\vskip 10mm

\noindent \underline{$K=3$}
\begin{center}
\begin{tabular}{|c||c|c|c|c|c||c|c|c|c|}\hline%\hline
  $\nr$&$\alpha(-1)$&$\alpha(0)$&$\alpha(1)$&$\alpha(2)$
  &$\alpha(3)$&$\beta(0)$&$\beta(1)$&$\beta(2)$&$\beta(3)$\\
  \hline
  1&0&1&0&0&0&2&0&0&0\\
  2&0&5&1&0&0&12&2&0&0\\
  3&0&19&5&1&0&50&10&2&0\\
  \hline
\end{tabular}
\end{center}
%\vskip 10mm

\noindent \underline{$K=4$}
\begin{center}
\begin{tabular}{|c||c|c|c|c|c||c|c|c|c|}\hline%\hline
  $\nr$&$\alpha(-1)$&$\alpha(0)$&$\alpha(1)$&$\alpha(2)$
  &$\alpha(3)$&$\beta(0)$&$\beta(1)$&$\beta(2)$&$\beta(3)$\\
  \hline
  1&0&1&0&0&0&2&0&0&0\\
  2&0&6&1&0&0&14&2&0&0\\
  3&0&27&6&1&0&68&12&2&0\\
  \hline
\end{tabular}
\end{center}

\noindent \underline{$K=5$}
\begin{center}
\begin{tabular}{|c||c|c|c|c|c||c|c|c|c|}\hline%\hline
  $\nr$&$\alpha(-1)$&$\alpha(0)$&$\alpha(1)$&$\alpha(2)$
  &$\alpha(3)$&$\beta(0)$&$\beta(1)$&$\beta(2)$&$\beta(3)$\\
  \hline
  1&0&1&0&0&0&2&0&0&0\\
  2&0&7&1&0&0&16&2&0&0\\
  \hline
\end{tabular}
\end{center}

%%%%%%%%%%%%%%%%%%%%%%%%%%%%%%%%%%%%%%%%%%%%
%%                                        %%
%%   Appendix B                           %%
%%                                        %%
%%%%%%%%%%%%%%%%%%%%%%%%%%%%%%%%%%%%%%%%%%%%
\section* {Appendix B: Free-fermion representation of
characters for the permutation and the general linear groups}
\setcounter{section}{2}
\setcounter{equation}{0}

Characters of the permutation group
and the general linear group can
be expressed in terms of free fermions \cite{rS}\cite{rDJKM}.
In this appendix we summarize the useful formulae.

%%%%%%%%%%%%%%%%%%%
% B.1             %
%%%%%%%%%%%%%%%%%%%
\subsection{}
{}Free fermions\footnote{
We use this notation to avoid a
confusion with the free fermions used
in the free-field realization of $W_{1+\infty}$.
Relation to usual free fermions
 $\bar{\psi}(z)=\sum_{r\in\bZ+1/2}\bar{\psi}_rz^{-r-1/2}$,
 $\psi(z)=\sum_{r\in{\bf Z}+1/2}\psi_rz^{-r-1/2}$ is given by
 $\fb_n=\bar{\psi}_{n+1/2}$, $\f_n=\psi_{n-1/2}$.
}
$\fb(z)=\sum_{n\in\bZ}\fb_nz^{-n-1}$,
$\f(z)=\sum_{n\in\bZ}\f_nz^{-n}$
and the vacuum state $\dket{0}$ are
defined by
\ba
  && \lbrace \fb_m, \f_n \rbrace = \delta_{m+n,0},\quad
  \lbrace \fb_m, \fb_n \rbrace =
\lbrace \f_m, \f_n \rbrace = 0, \n
  && \fb_m \dket{0} = \f_n \dket{0} =0, \qquad (m\geq 0, n\geq 1).
\ea
The fermion Fock space is a linear span of
$\prod_i \fb_{-m_i} \prod_j \f_{-n_j} \dket{0}$.
The $U(1)$ current $\Jc(z)=
\sum_{n\in\bZ}\Jc_nz^{-n-1}$ is defined by
$\Jc(z)=:\fb(z)\f(z):$, {\it i.e.},
$\Jc_n=\sum_{m\in\bZ}:\fb_m\f_{n-m}:$,
where the normal ordering $:\fb_m\f_n:$ means
$\fb_m\f_n$ if $m\leq -1$ and $-\f_n\fb_m$ if $m\geq 0$.
Their commutation relations are
\be
  \lbrack \Jc_n, \Jc_m \rbrack = n\delta_{n+m,0},\quad
  \lbrack \Jc_n, \fb_m \rbrack = \fb_{n+m},\quad
  \lbrack \Jc_n, \f_m \rbrack = -\f_{n+m}.
\ee

To the Young diagram $Y$ of the following form
($m_1>\cdots>m_h\geq 1$, $n_1>\cdots>n_h \geq 0$):
%%%%%%%%%%%%%%%%%%%%%%
%                    %
%  Figure            %
%                    %
%%%%%%%%%%%%%%%%%%%%%%

\centerline{\epsfbox{fig2.eps}}
%figfigfigfigfigfigfigfig\\
%\centerline{Hook decomposition of Young diagram $Y$.}

%%%%%%%%%%%%%%%%%%%%%%
%                    %
%  End of Figure     %
%                    %
%%%%%%%%%%%%%%%%%%%%%%
\noindent
we define the corresponding state $\dket{Y}$ as
\be
  \dket{Y}\equiv\prod_{i=1}^h
\fb_{-m_i}\f_{-n_i} (-1)^{n_i} \dket{0}.
  \label{Y}
\ee
Note that the number of fermion bilinears,
$h$, corresponds to that of
hooks in the Young diagram.
Bra states are obtained from ket states by $\dagger$ operation
($\fb_n{}^{\dagger}=\f_{-n}$) with the normalization
$\langle\!\langle0\dket{0}=1$; for example,
$\dbra{Y}=\dket{Y}^{\dagger}=\dbra{0}
\prod_{i=1}^h \fb_{n_i}\f_{m_i} (-1)^{n_i}$ and
$\langle\!\langle Y\dket{Y'}=\delta_{YY'}$.
Note that $\{ \dket{Y} \}$ is an
orthonormal basis of the fermion Fock
space with vanishing $U(1)$ charge.

Irreducible representations of the permutation group $\Sn$ and the
general linear group $GL(N)$ are both characterized by the Young
diagrams $Y$.
We denote their characters by
$\chi_Y(k)$ and $\tau_Y(x)$, respectively.
Here $(k)=1^{k_1}2^{k_2}\cdots
n^{k_n}$ stands for the conjugacy class
of $\Sn$; $k_1+2k_2+\cdots+nk_n=n=$the number of boxes in $Y$.
$x=[x_{\ell}]$ ($\ell=1,2,3,\cdots$) stands for
$x_{\ell}=\frac{1}{\ell}\mbox{tr }g^{\ell}
=\frac{1}{\ell}\sum_{i=1}^N
\epsilon_i^{\ell}$ for an element $g$ of
$GL(N)$, and in this case the number of boxes in $Y$ is a rank of
tensor for $GL(N)$.
In section 6, we consider the Lie
algebra of $GL(N)$, $\mbox{gl}(N)$,
for sufficiently large $N$.

$\chi_Y(k)$ and $\tau_Y(x)$ are expressed as follows:
\ba
  \chi_Y(k) &\!\!=\!\!&
  \dbra{0} \Jc_1^{k_1}\Jc_2^{k_2}\cdots\Jc_n^{k_n} \dket{Y},
  \label{chiY} \\
  \tau_Y(x) &\!\!=\!\!&
  \dbra{0}\exp\biggl(\sum_{\ell=1}^{\infty}x_{\ell}\Jc_{\ell}\biggr)
  \dket{Y}.
\ea
We remark that they can also be written as
$\chi_Y(k)= \dbra{Y}
\Jc_{-1}^{k_1}\Jc_{-2}^{k_2}\cdots\Jc_{-n}^{k_n} \dket{0}$ and
$\tau_Y(x)=\dbra{Y} \exp \left(
\sum_{\ell=1}^{\infty} x_{\ell}\Jc_{-\ell} \right) \dket{0}$.

%%%%%%%%%%%%%%%%%%%
% B.2             %
%%%%%%%%%%%%%%%%%%%
\subsection{}
{}For arbitrary parameters $u_r$ and $v_s$ such that the following
(infinite) product converges,
we can show the following identity,
\be
  \prod_r\prod_s \frac{1}{1-u_rv_s}
  =
  \sum_Y \tau_Y(x)\tau_Y(y),
  \label{YY}
\ee
where the summation runs over all the Young diagrams and $x,y$ are
the Miwa variables for $u,v$,
\be
  x_{\ell}\equiv\frac{1}{\ell}\sum_ru_r^{\ell},\quad
  y_{\ell}\equiv\frac{1}{\ell}\sum_sv_s^{\ell},\quad
  (\ell=1,2,3,\cdots).
\ee
\noindent{\bf Proof:}%(Proof)
\begin{eqnarray*}
  \prod_r\prod_s \frac{1}{1-u_rv_s}
  &\!\!=\!\!&
  \exp\biggl(\sum_r\sum_s\log\frac{1}{1-u_rv_s}\biggr)=
  \exp\biggl(\sum_r\sum_s
  \sum_{\ell=1}^{\infty}\frac{1}{\ell}(u_rv_s)^{\ell}\biggr) \\
  &\!\!=\!\!&
  \exp\biggl(\sum_{\ell=1}^{\infty}\ell x_{\ell}y_{\ell}\biggr) \\
  &\!\!=\!\!&
  \dbra{0}\exp\biggl(\sum_{\ell=1}^{\infty}
x_{\ell}\Jc_{\ell}\biggr)
  \exp\biggl(\sum_{\ell=1}^{\infty}
y_{\ell}\Jc_{-\ell}\biggr)\dket{0} \\
  &\!\!=\!\!&
  \sum_Y
  \dbra{0}\exp\biggl(\sum_{\ell=1}^{\infty}
x_{\ell}\Jc_{\ell}\biggr)
  \dket{Y}\dbra{Y}
  \exp\biggl(\sum_{\ell=1}^{\infty}y_{\ell}
\Jc_{-\ell}\biggr)\dket{0} \\
  &\!\!=\!\!&
  \sum_Y\tau_Y(x)\tau_Y(y).
\end{eqnarray*}
We have used the completeness of
$\{ \dket{Y} \}$ in the fermion Fock
space with vanishing $U(1)$ charge. \qed %\hfill $\Box$

%%%%%%%%%%%%%%%%%%%
% B.3             %
%%%%%%%%%%%%%%%%%%%
\subsection{}
In subsection 5.3 we need the following quantity for $\Sn$,
\be
  a^Y_n\equiv\frac{(-1)^n}{n!}\sum_{(k)}(-C)^{L(k)}N(k)\chi_Y(k),
\ee
where $(k)=1^{k_1}2^{k_2}\cdots n^{k_n}$,
$k_1+2k_2+\cdots+nk_n=n$, $L(k)=k_1+k_2+\cdots+k_n$ and
$N(k)=n!/(1^{k_1}k_1!2^{k_2}k_2!\cdots n^{k_n}k_n!)$.
We remark that $a^Y_n$ is a polynomial of $C$ with degree $n$,
$$
  a^Y_n=\frac{d_Y}{n!}C^n+\cdots.
$$

To calculate $a^Y_n$, we introduce its generating function
$a^Y(t)=\sum_{n=0}^{\infty}a^Y_nt^n$.
By eq.\ (\ref{chiY}), $a^Y(t)$ becomes
\be
  a^Y(t)=\dbra{0}\exp\biggl(C\sum_{\ell=1}^{\infty}
  \frac{(-1)^{\ell-1}}{\ell}\Jc_{\ell}t^{\ell}\biggr)\dket{Y}.
\ee
By rewriting eq.\ (\ref{Y}) as
\begin{eqnarray*}
  \dket{Y} &\!\!=\!\!&
  \prod_{i=1}^h(-1)^{n_i+i-1}\cdot
  \fb_{-m_1}\cdots\fb_{-m_h}\f_{-n_1}\cdots\f_{-n_h}\dket{0} \\
  &\!\!=\!\!&
  \oint \prod_{i=1}^{h}\frac{dz_i}{2\pi i}\frac{dw_i}{2\pi i}
  \prod_{i=1}^hz_i^{-m_i}w_i^{-n_i-1} \\
  &&~~~\times\,
  \prod_{i=1}^h(-1)^{n_i+i-1}\cdot
  \fb(z_1)\cdots\fb(z_h)\f(w_1)\cdots\f(w_h)\dket{0},
\end{eqnarray*}
$a^Y(t)$ can be expressed as
\begin{eqnarray*}
  a^Y(t) &\!\!=\!\!&
  \oint \prod_{i=1}^{h}\frac{dz_i}{2\pi i}\frac{dw_i}{2\pi i}
  \prod_{i=1}^hz_i^{-m_i}w_i^{-n_i-1} \\
  &&~~~\times\,
  \prod_{i=1}^h\frac{(1+tz_i)^C}{(1+tw_i)^C}
  (-1)^{n_i+i-1}\cdot
  \dbra{0}\fb(z_1)\cdots\fb(z_h)\f(w_1)\cdots\f(w_h)\dket{0}.
\end{eqnarray*}
Here we have used
\ba
  \exp\biggl(\sum_{\ell=1}^{\infty}x_{\ell}\Jc_{\ell}\biggr)
  \fb(z)
  \exp\biggl(-\sum_{\ell=1}^{\infty}x_{\ell}\Jc_{\ell}\biggr)
  &\!\!=\!\!&
  \exp\biggl(\sum_{\ell=1}^{\infty}x_{\ell}z^{\ell}\biggr)
  \fb(z), \n
  \exp\biggl(\sum_{\ell=1}^{\infty}x_{\ell}\Jc_{\ell}\biggr)
  \f(z)
  \exp\biggl(-\sum_{\ell=1}^{\infty}x_{\ell}\Jc_{\ell}\biggr)
  &\!\!=\!\!&
  \exp\biggl(-\sum_{\ell=1}^{\infty}x_{\ell}z^{\ell}\biggr)
  \f(z),
  \label{JfJ}
\ea
in particular, for
$U(C)=\exp\Bigl(C\sum_{\ell=1}^{\infty}
\frac{(-1)^{\ell-1}}{\ell}\Jc_{\ell}t^{\ell}\Bigr)$
\begin{eqnarray*}
  U(C)\fb(z)U(-C)&\!\!=\!\!&(1+tz)^C \fb(z), \n
  U(C)\f(z)U(-C)&\!\!=\!\!&(1+tz)^{-C} \f(z).
\end{eqnarray*}
By expanding $a^Y(t)$, we obtain
$$
  a^Y_n=
  \sum_{r_i,s_i}
  \prod_{i=1}^h {C \choose m_i+r_i}{-C \choose n_i-s_i}
  (-1)^{n_i+i-1}\cdot
  \dbra{0}\fb_{r_1}\cdots\fb_{r_h}\f_{-s_1}\cdots\f_{-s_h}\dket{0},
$$
where the summation runs over $r_i\geq 0$, $0\leq s_i\leq n_i$,
$\sum_{i=1}^hr_i=\sum_{i=1}^hs_i$. ${x \choose n}$ is defined by
${x \choose n}=[x]_n/n!$ and $[x]_n=\prod_{i=0}^{n-1}(x-i)$.
Thus, $a^Y_n$ is divided by $\prod_{i=1}^h{C \choose m_i}$
as a polynomial of $C$.
Similarly, starting from
\begin{eqnarray*}
  \dket{Y}&\!\!=\!\!&
  \oint \prod_{i=1}^{h}\frac{dz_i}{2\pi i}\frac{dw_i}{2\pi i}
  \prod_{i=1}^hz_i^{-m_i}w_i^{-n_i-1} \\
  &&~~~\times\,
  \prod_{i=1}^h(-1)^{n_i+i-1+h}\cdot
  \f(w_1)\cdots\f(w_h)\fb(z_1)\cdots\fb(z_h)\dket{0},
\end{eqnarray*}
we obtain
$$
  a^Y_n=
  \sum_{r_i,s_i}
  \prod_{i=1}^h {C \choose m_i-r_i}{-C \choose n_i+s_i}
  (-1)^{n_i+i-1+h}\cdot
  \dbra{0}\f_{s_1}\cdots\f_{s_h}\fb_{-r_1}\cdots\fb_{-r_h}\dket{0},
$$
where the summation runs over $1\leq r_i\leq m_i$, $s_i\geq 1$,
$\sum_{i=1}^hr_i=\sum_{i=1}^hs_i$.
Therefore, $a^Y_n$ is divided by $\prod_{i=1}^h{-C \choose n_i+1}$
as a polynomial of $C$.
Combining these results, we can conclude that $a^Y_n$ is
$\prod_{i=1}^h{C \choose m_i}{-C \choose n_i+1}/C^h$ up to constant.
We thus finally obtain
\be
  a^Y_n=\frac{d_Y}{n!}\prod_{i=1}^h[C]_{m_i}[-C-1]_{n_i}(-1)^{n_i}.
\ee
This result can be converted into a
simpler form as given in subsection
5.3:
\be
  a^Y_n=\frac{d_Y}{n!}\prod_{b\in Y}(C-C_b).
\ee

%% ===============================================================
%%% original proof %%%
We can give another easy proof of the above result.
The relation between the transformed basis
$U(C)\fb_{-m}U(-C)$ with the original ones
$\fb_{-m}$ can be obtained if we expand the factor
$(1+tz)^C$ in $z$ around $0$.
{}For non-integer $C$, $U(C)\fb_{-m}U(-C)$ is written by
infinite sum with respect to $\fb_{-m+\ell}$ with
$\ell=0,1,2,3,\cdots$.  However, if $C\in \ZZ$,
truncation of the summation happens.
%\ba
%U(C)\flat_n U(-C)&=& \sum_{\ell=0}^{C} \flat_{n+\ell}
%\qquad \mbox{if }C>0,\n
%U(C)\flatd_n U(-C)&=& \sum_{\ell=0}^{|C|} \flatd_{n+\ell}
%\qquad \mbox{if }C<0.
%\label{d25}
%\ea
%
We move each operator $\fb_{-m}\f_{-n}$
in \eq{Y} to
the left of $U(C)$.  When it is acted on the bra vacuum,
it vanishes when $C=-n,-n+1,\cdots,m-2,m-1$.
It gives the following assignment
of the polynomial of $C$ to each pair of the fermion operators:
\be
\fb_{-m}\flat_{-n}
\Longleftrightarrow
(C+n)(C+n-1)\dots(C-m+2)(C-m+1).
\label{d26}
\ee
Combining these factors for each hook, we get the
assignment in \eq{I11}.

%%%%%%%%%%%%%%%%%%%%%%%%%%%%%%%%%%%%%%%%%%%%%%%%%%%%%%%%
%
%         references
%
%%%%%%%%%%%%%%%%%%%%%%%%%%%%%%%%%%%%%%%%%%%%%%%%%%%%%%%%%%

\end{document}